\documentclass[final,5p,twocolumn,number,10pt]{elsarticle}

\usepackage{amssymb}

\usepackage{amsmath}
\usepackage{color}
\usepackage{here}
\usepackage{bm}
\biboptions{sort&compress}

\begin{document}

\begin{frontmatter}

\title{Constraints on the nuclear Schiff moment from its correlation with electromagnetic properties}

\author[1]{K.~Yanase}

\affiliation[1]{
  organization={Center for Nuclear Study, The University of Tokyo},
  addressline={Hongo 7-3-1},
  postcode={113-0033},
  city={Bunkyo-ku},
  state={Tokyo},
  country={Japan}
}
\ead{yanase@cns.s.u-tokyo.ac.jp}

\author[2]{N.~Shimizu}

\affiliation[2]{
  organization={Center for Computational Sciences, University of Tsukuba},
  city={Tsukuba},
  postcode={305-8577},
  country={Japan}
}
\ead{shimizu@nucl.ph.tsukuba.ac.jp}

\author[3]{K.~Higashiyama}

\affiliation[3]{
  organization={Department of Physics, Chiba Institute of Technology},
  city={Narashino},
  postcode={275-0023},
  state={Chiba},
  country={Japan}
}

\author[4]{N.~Yoshinaga}
\affiliation[4]{
  organization={Department of Physics, Saitama University},
  city={Saitama City},
  postcode={338-8570},
  country={Japan}
}

\begin{abstract}
We study the nuclear Schiff moments of $^{129}$Xe and $^{199}$Hg induced by the nucleon electric dipole moment using large-scale shell model calculations.
For $^{129}$Xe, we find a linear relation between the leading-order contribution and magnetic moment, which would be useful in reducing the theoretical uncertainty.
The conventional model space does not contain the $ 0g_{ 9/2 } $ and $ 0h_{ 9/2 } $ orbitals, which are connected to the spin-orbit partners by large matrix elements.
Thus, to evaluate the influence of the relevant single-particle orbitals outside the conventional model space, we apply the quasiparticle vacua shell model method.
Moreover, the next-to-leading-order contribution arises from parity and time-reversal violation inside the nucleus.
We demonstrate that these secondary effects do not induce any significant disturbance to the correlation.
Additionally, we report the shell model results for the nuclear Schiff moment coefficients of $^{199}$Hg.
Compared to previous studies, the results obtained in this study are rather large, indicating a higher sensitivity to the neutron electric dipole moment.
\end{abstract}


\end{frontmatter}

\section{Introduction}
\label{sec_intro}

Charge-parity ($ CP $) violation in fundamental physics is one of the necessary conditions for the matter-antimatter asymmetry in the current universe~\cite{Trodden1999-review,Dine2003-review,Bodeker2021-review}.
The permanent electric dipole moments (EDMs) of elementary and composite particles are promising probes for $ CP $ violations beyond the standard model.
In particular, atomic EDM is nonzero only if parity ($ P $) and time-reversal ($ T $) symmetries are violated by the interactions between the constituent particles of an atom.

Several $ P $, $ T $-odd coupling constants may exist at the nuclear energy scale~\cite{Engel2013-review,Yamanaka2017-review,Chupp2019-review}.
Recent measurements of paramagnetic atoms and molecules have placed upper limits on the electron EDM~\cite{Regan2002,ACME2014,ACME2018}.
In contrast, atomic EDM induced by the electron EDM is suppressed for diamagnetic atoms forming a closed-shell electronic structure~\cite{Ginges2004-review,Maartensson1987,Dzuba2009-diamagnetic}.
While this implies an insensitivity to the electron EDM, the diamagnetic atoms represented by $^{129}$Xe and $^{199}$Hg can provide background-free searches for hadronic $ P $ and $ T $ violations.

The basic concept of the atomic EDM experiments involves precisely measuring the energy shift of neutral atoms exposed to an applied electric field.
In principle, the EDMs of constituent particles can be coupled to an external electric field and contribute to the energy shift.
However, Schiff's theorem states that these contributions are completely screened in the non-relativistic and point-like nucleus limits~\cite{Schiff1963,Liu2007-nEDM,Liu2007-NSM,Senkov2008,Yanase2021-screening}.
Fortunately, this screening phenomenon can be avoided in finite-size nuclei, which implies that heavy nuclei may be suitable for atomic EDM measurements to explore hadronic $ P $ and $ T $ violations.
The detectable $ P $, $ T $-odd nuclear moments are derived from the multipole expansion of the nuclear distribution and the leading part is referred to as the nuclear Schiff moment (NSM).
The NSMs of diamagnetic atoms have not been calculated with high precision to date.
In this study, we compute the NSMs of $^{129}$Xe and $^{199}$Hg induced by the nucleon EDM using the nuclear shell model.

\section{Nuclear Schiff moment}
\label{sec_nsm}

The NSM is factorized into $ P $, $ T $-odd coupling constants at the nuclear energy scale and their coefficients, $ s_{ N } $, $ a_{ T } $, and $ a_{ N } $, obtained by nuclear many-body calculations as
\begin{align}
 &
 S = S_{ 2 } + S_{ 3 }
 \\
 &
 S_{ 2 }
 =
 \sum_{ N = p, n }
 s_{ N } d_{ N }
 ,
 \\
 &
 S_{ 3 }
 =
 \sum_{ T = 0 }^{ 2 }
 a_{ T }
 g_{ \pi NN }^{( T )}
 +
 \sum_{ N = p, n }
 a_{ N }
 d_{ N }
 ,
\end{align}
where $ d_{ p } $ and $ d_{ n } $ are the proton and neutron EDMs, respectively.
The second-order contribution, $ S_{ 2 } $, arises from the the direct $ P $ and $ T $ violations of the electron system through the interactions between the nucleon EDMs and electrons.
In addition, the $ P $, $ T $-odd nucleon-nucleon ($ NN $) interactions induce nuclear EDM violating the $ P $ and $ T $ symmetries of the electron system.
This is a third-order process, and its contribution to the NSM is denoted by $ S_{ 3 } $.
Here, $ g_{ \pi NN }^{( T )} $ denotes the $ P $, $ T $-odd coupling constants of the one-pion-exchange nucleon-nucleon ($ \pi NN $) interactions with the isospin components $ T = 0, 1, 2 $~\cite{Herczeg1988}.
The NSM coefficients $ a_{ T } $, induced by the $ P $, $ T $-odd $ \pi NN $ interactions, were computed based on large-scale shell model calculations~\cite{Yanase2020-129Xe-199Hg}.
Here, we focus on $ s_{ N } $ and $ a_{ N } $ induced by the nucleon EDM.

The second-order NSM operator is defined by
\begin{align}
 S_{ 2, k }
 & =
 \frac{ 1 }{ 6 }
 \sum_{ a = 1 }^{ A }
 d_{ a, k }
 \big(
  r_{ a }^{ 2 }
  -
  \big\langle r^{ 2 } \big\rangle_{ \text{ch} }
 \big)
 \notag\\
 & \quad
 +
 \frac{ 2 }{ 15 }
 \sum_{ a = 1 }^{ A }
 \sum_{ j }
 d_{ a, j }
 \big(
  Q_{ a, jk }
  -
  \big\langle Q_{ jk } \big\rangle_{ \text{ch} }
 \big)
 \label{eq: S2 operator}
 ,
\end{align}
where $ \sigma_{ k } $ is the spin Pauli matrix, and $ Q_{ a, jk } $ is the nuclear quadrupole moment.
The nucleon EDMs are represented by $ d_{ a, k } = d_{ p } \sigma_{ k } $ for protons and $ d_{ a, k } = d_{ n } \sigma_{ k } $ for neutrons.
The cartesian components are denoted by $ j $ and $ k $.
The second-order contribution to the NSM is computed by
\begin{align}
 S_{ 2 }
 =
 \big\langle \psi_{ \text{g.s.} } \big|
  S_{ 2z }
 \big| \psi_{ \text{g.s.} } \big\rangle
 \label{eq: S2 expectation value}
 ,
\end{align}
where $ \big| \psi_{ \text{g.s.} } \big\rangle $ represents the nuclear ground state with definite parity.
The $ P $ and $ T $ symmetries are not violated in the ground state.
In contrast, the third-order processes involve $ P $ and $ T $ violations inside the nucleus.
In fact, the third-order contribution is given by the $ P $, $ T $-odd operator
\begin{align}
 \bm{ S }_{ 3 }
 =
 \frac{ e }{ 10 }
 \sum_{ a = 1 }^{ Z }
 \bigg[
  r_{ a }^{ 2 }
  \bm{ r }_{ a }
  -
  \frac{ 5 }{ 3 }
  \bm{ r }_{ a }
  \left\langle r^{ 2 } \right\rangle_{ \text{ch} }
  -
  \frac{ 4 }{ 3 }
  \bm{ r }_{ a }
  \left\langle Q_{ jk } \right\rangle_{ \text{ch} }
 \bigg]
 \label{eq: S3 operator}
 ,
\end{align}
and the expectation value is calculated perturbatively as
\begin{align}
 S_{ 3 }
 =
 &
 \sum_{ n }
 \frac{ 1 }{ E_{ \text{g.s.} } - E_{ n } }
 \big\langle \psi_{ \text{g.s.} } \big|
  S_{ 3z }
 \big| \psi_{ n } \big\rangle
 \big\langle \psi_{ n } \big|
  \widetilde{ V }
 \big| \psi_{ \text{g.s.} } \big\rangle
 \notag\\
 &
 + c.c.
 \label{eq: S3 expectation value}
\end{align}
where $ \widetilde{ V } $ denotes the $ P $, $ T $-odd $ NN $ interactions.

Theoretical studies on $ a_{ T } $ of $^{199}$Hg are of particular interest because of the current best limits on the $ P $, $ T $-odd $ \pi NN $ coupling constants provided by recent $^{199}$Hg atomic EDM experiments~\cite{Graner2016,Graner2017-erratum}.
The fully self-consistent Hartree-Fock (HF) and Hartree-Fock Bogoliubov (HFB) calculations predict a significant reduction in $ a_{ T } $ from the simple estimate of the independent particle model (IPM) owing to core polarization~\cite{Ban2010}.
In contrast, a recent study reported a $ 12 \% $ reduction factor at most, based on the large-scale shell model (LSSM) calculation~\cite{Yanase2020-129Xe-199Hg}.
This discrepancy might be attributed to the destructive interference between neutrons and protons, nuclear deformation, and contamination by certain excited states where the NSM coefficients $ a_{ T } $ are orders of magnitude smaller than the ground state~\cite{Jesus2005,Ban2010,Yanase2020-129Xe-199Hg}.
The LSSM calculation shows that the ground state can be distinguished from the second lowest $ \tfrac{ 1 }{ 2 }^- $ state by spectroscopic factors and $ E2 $ transition strengths from the collective excited states.

The second-order NSM coefficients $ s_{ N } $ were evaluated for $^{129}$Xe and $^{199}$Hg, where certain discrepancies were found.
The sensitivity to the neutron EDM, $ s_{ n } $, is strongly reduced relative to the IPM estimate in the HF and HFB calculations for $^{199}$Hg~\cite{Ban2010}, whereas another mean-field-based calculation shows a moderate reduction~\cite{Dmitriev2003-PRL}.
For $^{129}$Xe, a pioneering shell model calculation of $ s_{ N } $ was performed using the nucleon-pair truncation scheme to reduce computational costs~\cite{Yoshinaga2010-NSM}.
Unexpectedly, this study predicted an opposite sign of $ s_{ n } $ with respect to the IPM estimate.
However, little attention has been paid to this problem, which can affect the constraints on the $ P $, $ T $-odd couplings at the fundamental level.
In particular, the contribution of the Weinberg operator is potentially enhanced, depending on the sign of $ s_{ n } $~\cite{Yamanaka2017-review,Osamura2022}.
Moreover, the nucleon EDM is the only probe of the quark EDM.

\section{Nuclear shell model}
\label{sec_shell-model}

\subsection{Large-scale shell-model calculations}
\label{subsec_LSSM}

In this study, we apply the LSSM calculations performed with the effective interactions SNV~\cite{Utsuno2014} and SN100PN~\cite{Brown2005} in order to compute $ s_{ N } $ and the third-order contribution induced by the nucleon EDM, $ a_{ N } $, for $^{129}$Xe.
Recent studies on nuclear spectroscopy have shown that low-energy spectra are reproduced by the LSSM calculations involving these realistic interactions in a specific area of the nuclear chart~\cite{
Kaya2018-131Xe,Kaya2018-133Xe-135Ba,
Kaya2019-136Ba-137Ba,Kaya2019-133Ba-134Ba,
Laskar2019-135La,Laskar2020-133La}.

\subsection{Quasiparticle vacua shell model}
\label{subsec_QVSM}

The effective one-body operator of the nuclear magnetic moment is expressed as
\begin{align}
 \bm{ \mu }
 =
 \sum_{ a = 1 }^{ A }
 \big(
  g_{ l, \text{eff} } \bm{ l }_{ a }
  +
  g_{ s, \text{eff} } \bm{ s }_{ a }
 \big)
 \label{eq: magnetic moment operator}
 ,
\end{align}
where $ \bm{ l }_{ a } $ and $ \bm{ s }_{ a } $ denote the orbital and spin angular momenta of the nucleon, respectively.
We adopt a quenching factor of $ 0.7 $ for the spin $ g $ factor, which is attributed to the two-body current correction and core polarization.
This value is fixed to reproduce the magnetic moments of $^{129,131,133,135}$Xe, which is consistent with other studies~\cite{Kaya2018-133Xe-135Ba,Rodriguez2020-133Sn}.
The LSSM calculations of $^{129}$Xe employ the standard model space between the magic numbers 50 and 82 and then lack the particle-hole configurations, which have large one-body matrix elements of the magnetic moment, such as $ ( 0g_{ 9/2 }^{ -1 }, 0g_{ 7/2 } ) $ and $ ( 0h_{ 11/2 }^{ -1 }, 0h_{ 9/2 } ) $.
The spin dependence of the second-order NSM operator in Eq.~(\ref{eq: S2 operator}) indicates that the one-body matrix elements between the spin-orbit partners are also significant.
We then apply the quasiparticle vacua shell model (QVSM) with the model space extended to include the $ 0g_{ 9/2 } $ and $ 0h_{ 9/2 } $ orbitals.
The QVSM is a variational approach to describe nuclear wave functions as linear combinations of the proton and neutron numbers, parity, and angular momentum projected basis states.
For odd-mass nuclei, the basis states are given by one-quasiparticle states characterized by variational parameters, which are determined to minimize the energy of the projected and superposed wave functions.
The framework of the QVSM calculation is detailed in Ref.~\cite{Shimizu2021-QVSM}.

\begin{table}[htb]
\caption{
\label{table: QVSM spectra}
Excitation energies of the lowest $ 2^+ $ and $ 4^+ $ states in units of MeV obtained by the QVSM calculations.
}
\begin{center}
\begin{tabular*}{0.48\textwidth}{@{\extracolsep{\fill}}cccccc} \hline
	&
			& $^{134}$Te 	& $^{130}$Sn 
			& $^{128}$Xe 	& $^{130}$Xe
			\\\hline
	$ E_{ x } ( 2^+_1 ) $
	& Expt.
			& $ 1.28 $ 		& $ 1.22 $
			& $ 0.44 $ 		& $ 0.54 $
			\\
	& SNV-$ sdgh $
			& $ 1.30 $ 		& $ 1.21 $
			& $ 0.42 $ 		& $ 0.57 $
			\\
	& SNJKV
			& $ 1.31 $ 		& $ 1.23 $
			& $ 0.43 $ 		& $ 0.63 $
			\\\hline
	$ E_{ x } ( 4^+_1 ) $
	& Expt.
			& $ 1.58 $ 		& $ 2.00 $
			& $ 1.03 $ 		& $ 1.21 $
			\\
	& SNV-$ sdgh $
			& $ 1.64 $ 		& $ 2.15 $
			& $ 1.14 $ 		& $ 1.42 $
			\\
	& SNJKV
			& $ 1.64 $ 		& $ 2.19 $
			& $ 1.16 $ 		& $ 1.52 $
			\\\hline
\end{tabular*}
\end{center}
\end{table}

We adopt phenomenological extensions of the SNV effective interaction with the full $ sdgh $ shell for the QVSM calculation.
The SNV effective interaction is employed for the standard model space, and the extended part is given by the VMU interaction~\cite{Utsuno2014}.
This interaction is referred to as SNV-$ sdgh $.
Another set, SNJKV interaction, consists of the SNV, JUN45, and Kuo-Herling interactions, and the remaining proton-neutron part is given by the VMU interaction.
To compensate for the enhancement in the pairing correlation by extending the model space, we reduce the pairing interaction strengths between like nucleons.
The reduction factors are determined by low-energy spectra of $^{134}$Sn and $^{134}$Te as 0.72 for proton and 0.76 for neutron in the SNV-$ sdgh $ interaction, and 0.74 for proton and 0.76 for neutron in the SNJKV interaction.

Table \ref{table: QVSM spectra} shows the excitation energies of the $ 2^+ $ and $ 4^+ $ states obtained by the QVSM calculations.
While the $ 2^+ $ states are systematically reproduced, the $ 4^+ $ states are higher than the experimental values by 0.2~MeV on average.
Here, benchmark calculations, based on the energy variance extrapolation method, show slower convergence of high-spin states~\cite{Shimizu2021-QVSM}.
This implies that the excitation energies of the $ 4^+ $ states are overestimated in the present QVSM calculations.
A natural progression of this study is to perform a global calculation using such an extrapolation procedure to further improve the effective interaction.


\begin{figure}[htb]
\begin{center}
\includegraphics[width=0.9\linewidth]{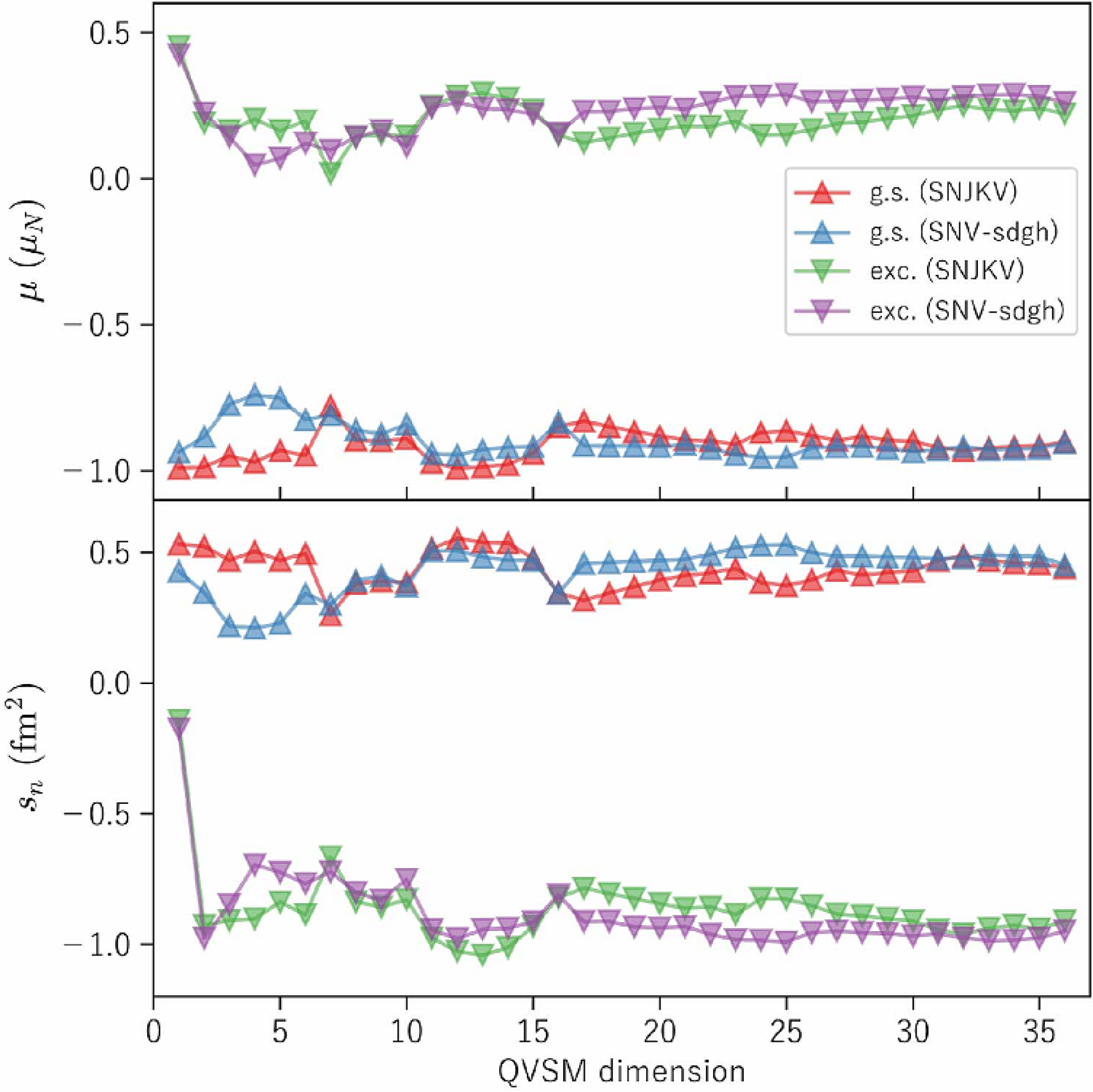}
\caption{
\label{fig: QVSM convergence}
Convergence of the magnetic moments $ \mu $ and second-order NSM coefficients $ s_{ n } $ of $^{129}$Xe.
The ground state (g.s.) and the second lowest $ 1/2^+ $ state (exc.) are represented by triangles and inverted triangles, respectively.
}
\end{center}
\end{figure}


Figure~\ref{fig: QVSM convergence} shows the convergence of the magnetic moment and second-order NSM coefficient $ s_{ n } $ as the number of QVSM basis states increases.
The standard deviation of the last 10 points is $ 5 \% $, at most.
In fact, this level of accuracy does not qualitatively change our conclusion.
Comparing the magnetic moment and NSM coefficient in the region of a small number of basis states, we find that they fluctuate in opposite directions.
This feature stems from the spin dependence of the operators in Eq.~(\ref{eq: S2 operator}) and (\ref{eq: magnetic moment operator}).
A similar behavior is found between the NSM coefficients of the ground state and the second-lowest $ 1/2^+ $ state.
As discussed below, these relations help reducing the uncertainty of the NSM coefficients.

\subsection{Pair-truncated shell model}
\label{subsec_PTSM}

The second-order NSM coefficients $ s_{ N } $ of $^{129}$Xe were investigated by using the pair-truncated shell model (PTSM), where many-body configurations are made of building blocks composed of nucleon pairs with specific angular momenta~\cite{Yoshinaga2010-NSM,Yoshinaga2013,Yoshinaga2014-NSM-erratum}.
This reduces the dimension of the many-body Hamiltonian matrix to be diagonalized by orders of magnitude in comparison with the conventional $ J $-scheme dimension.
The PTSM calculation of $ s_{ N } $ was originally performed in Ref.~\cite{Yoshinaga2010-NSM}.
The authors employed a phenomenological monopole and quadrupole pairing plus quadrupole-quadrupole interaction as an effective interaction.
The interaction strengths were tuned to reproduce the low-energy spectra of Xe, Ba, Ce, and Nd isotopes close to the neutron magic number.
Here, we refer to this effective interaction as YH04.
Another parameter set of the same form of the effective interaction (YH11) was suggested in Ref.~\cite{Higashiyama2011} and applied to compute the NSM coefficients~\cite{Yoshinaga2013}.
This revision was introduced for the better reproduction of odd-mass nuclei.

In this study, we revisit the PTSM calculations with certain modifications.
We adopt the standard formula $ \hbar \omega = 41 A^{ -1/3 } $~(MeV) for the harmonic oscillator frequency to evaluate the one-body matrix elements as well as in the LSSM and QVSM calculations.
This choice corresponds to a root-mean-square charge radius of $ 4.6 $~fm for $^{129}$Xe, which is consistent with the experimental value of $ 4.8 $~fm~\cite{Angeli2013}.
In contrast, an incorrect value of $ \hbar \omega = 41 $~MeV was adopted in Ref.~\cite{Higashiyama2011}, and the root-mean-square charge radius was severely underestimated.
Moreover, the original study in Ref.~\cite{Yoshinaga2013} adopted an extended model space for protons, whereas we found that the influence to be negligible.
Therefore, we employ the standard model space for consistency with the LSSM calculations.


\begin{table}[htb]
\caption{
\label{table: NSM of Xe-129}
Calculated second-order NSM coefficients of $^{129}$Xe in units of $ \text{fm}^{ 2 } $.
}
\begin{center}
\begin{tabular*}{0.48\textwidth}{@{\extracolsep{\fill}}lcc} \hline
		&	$ s_{ p } $		& $ s_{ n } $ \\\hline
	LSSM (SNV)
  		& $ - 0.002 $ 	& $ 0.42 $ \\
	LSSM (SN100PN)
  		& $ 0.009 $ 		& $ 0.42 $ \\
	QVSM (SNV-$ sdgh $)
  		& $ - 0.005 $ 	& $ 0.48 $ \\
	QVSM (SNJKV)
  		& $ - 0.008 $ 	& $ 0.45 $ \\[5pt]
	PTSM (YH04)
  		& $ 0.006 $ 		& $ - 0.32 $ \\
	PTSM (YH11)
  		& $ 0.009 $ 		& $ - 0.48 $ \\\hline
\end{tabular*}
\end{center}
\end{table}


Table~\ref{table: NSM of Xe-129} presents the results of the shell model calculations for the second-order NSM coefficients.
We adopt the mean values of the last 10 points in Fig.~\ref{fig: QVSM convergence} as results of the QVSM calculations.
It is confirmed that the NSM of the neutron-odd nucleus, $^{129}$Xe, is more sensitive to the neutron EDM, whereas the sensitivity to the proton EDM is weak and strongly dependent on the effective interaction.
However, the PTSM calculations predict the opposite sign of $ s_{ n } $ to the LSSM and QVSM calculations.
This infers a serious discrepancy, which will be investigated hereafter.

\section{Correlations between the nuclear Schiff moment coefficients and electromagnetic properties}
\label{sec_correlations}

\subsection{$^{129}$Xe}
\label{subsec_discussion-129Xe}

\begin{figure}[hbt]
\begin{center}
\includegraphics[width=0.9\linewidth]{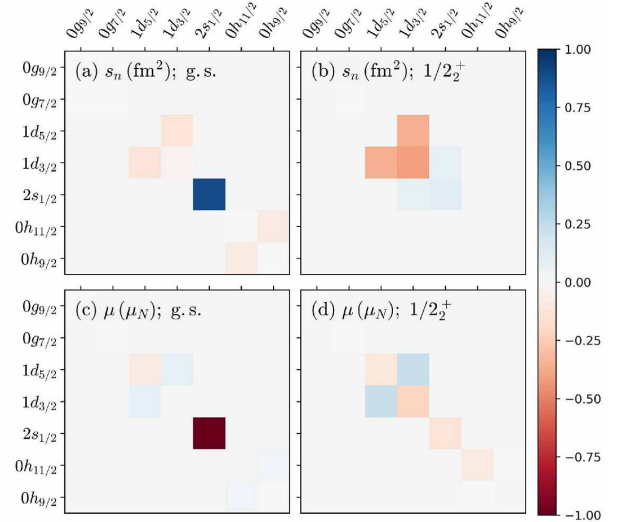}
\caption{
\label{fig: obtd_decomposition}
One-body transition density decomposition of the magnetic moment of (a) the ground state and (b) the second lowest $ 1/2^{ + } $ state.
Second-order NSM coefficient of (c) the ground state and (d) the second lowest $ 1/2^{ + } $ state.
}
\end{center}
\end{figure}

The second-order contribution is decomposed with respect to the one-body transition density as follows:
\begin{align}
 S_{ 2 }
 =
 \sum_{ ij }
 \big\langle i \big|
  S_{ 2z }
 \big| j \big\rangle
 \big\langle \psi_{ \text{g.s.} } \big|
  \big[
   c_{ i }^{ \dagger }
   \otimes
   \widetilde{ c }_{ j }
  \big]^{( 1 )}_{ 0 }
 \big| \psi_{ \text{g.s.} } \big\rangle
 \label{eq: S2 obtd}
 ,
\end{align}
where $ i, j $ indicate the single-particle orbitals.
Figure~\ref{fig: obtd_decomposition}~(a) shows each term obtained by the QVSM calculation using the SNJKV interaction.
The odd neutron occupies the $ 2s_{ 1/2 } $ orbital in the IPM, and the diagonal element of this orbital is still dominant in the QVSM calculation.
The spin-flipping contributions reduce $ s_{ n } $ by half with respect to the IPM estimate, while preserving the sign.
To investigate the significant decrease in the PTSM calculations, we performed the same analysis for the second-lowest $ 1/2^+ $ state.
As shown in Fig.~\ref{fig: obtd_decomposition}~(b), the single-particle nature of the $ 2s_{ 1/2 } $ orbital disappears, and the negative contributions of the $ d $ orbitals become dominant.


Figures~\ref{fig: obtd_decomposition}~(c) and (d) show the same decomposition of the magnetic moment.
One can find similarity with the NSM coefficient, which recalls the synchronized behavior shown in Fig.~\ref{fig: QVSM convergence}.
Considering that the second-order NSM operator~(\ref{eq: S2 operator}) is proportional to the spin operator, associating $ s_{ n } $ with the magnetic moment is reasonable.
In fact, the magnetic moment is dominated by the neutron spin part, because the odd neutron in the $ 2s_{ 1/2 } $ orbital has zero orbital angular momentum.

\begin{figure}[htb]
\begin{center}
\includegraphics[width=0.9\linewidth]{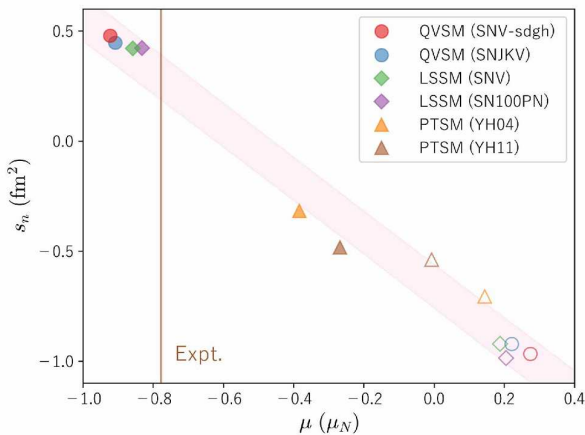}
\caption{
\label{fig: mag-sn correlation of Xe-129}
Correlation between the magnetic moments $ \mu $ and the second-order NSM coefficients $ s_{ n } $ of $^{129}$Xe.
The filled and open symbols represent the ground state and the second lowest $ 1/2^+ $ state, respectively.
}
\end{center}
\end{figure}

The possible correlation between the second-order NSM coefficient $ s_{ n } $ and the magnetic moment mentioned above is demonstrated more explicitly in Fig.~\ref{fig: mag-sn correlation of Xe-129}.
This correlation is understood by considering a two-level model that describes the ground state as
\begin{align}
 \big| \psi \big\rangle
 =
 \alpha
 \big| \psi_{ \text{g.s.} } \big\rangle
 +
 \sqrt{ 1 - \alpha^{ 2 } }
 \big| \psi_{ 1/2^+_2 } \big\rangle
 ,
\end{align}
where $ \big| \psi_{ \text{g.s.} } \big\rangle $ and $ \big| \psi_{ 1/2^+_2 } \big\rangle $ denote the ideal ground state that reproduces the experimental value of the magnetic moment, and the second lowest $ 1/2^+ $ state, respectively.
If a nuclear model calculation represents the ground state as $ \alpha \simeq 1 $, the result should be plotted on the upper left in Fig.~\ref{fig: mag-sn correlation of Xe-129} corresponding to the experimental value of the magnetic moment.
However, the result approaches the predicted values of the second-lowest $ 1/2^+ $ state when $ \alpha $ approaches zero.
This model can explain the strong quenching predicted by the PTSM calculations, which is corroborated by the isotopic dependence.
As shown in Fig.~2 in Ref.~\cite{Yoshinaga2010-NSM}, the NSM coefficient $ s_{ n } $ of $^{135}$Xe is close to the IPM estimate in the PTSM calculations with the YH04 interaction.
Because the $^{135}$Xe nucleus consists of one neutron hole in the $ 50 \leq N \leq 82 $ model space, the ground-state wave function is similar to the IPM wave function in shell model calculations with any truncation schemes and effective interactions.
However, the NSM coefficient $ s_{ n } $ is drastically reduced as the neutron number decreases and many-body correlations come into play.
Considering the inconsistency with the experimental value of the magnetic moment, the PTSM calculations would fail to describe the relatively complicated wave function of $^{129}$Xe, which consists of seven neutron holes in the conventional model space.
Consequently, the contamination of the ideal second-lowest $ 1/2^+ $ state causes a strong quenching of the NSM coefficient and magnetic moment.


Here, we mention the effect of extending the model space in the QVSM calculations.
As explained above, the $ 0h_{ 11/2 } $ orbital is strongly connected with its spin-orbit partner, the $ 0h_{ 9/2 } $ orbital, by the second-order NSM and magnetic moment operators.
The $ 0h_{ 9/2 } - 0h_{ 11/2 } $ contributions placed at the bottom right corners of Figs.~\ref{fig: obtd_decomposition}~(a) and (c) are opposite to the major components of the $ 2s_{ 1/2 } $ orbital.
However, the magnetic moment is larger than the results of the LSSM calculations without the $ 0h_{ 9/2 } $ orbital, possibly because the magnetic moment of $^{129}$Xe is sensitive to configuration mixing inside the conventional $ 50 \leq Z, N \leq 82 $ model space~\cite{Arima1954}.

Using the correlation with the magnetic moment, the theoretical uncertainty of the NSM coefficient $ s_{ n } $ can be considerably reduced.
Although the PTSM results are counted with equal weights, we obtain $ s_{ n } = 0.29 \pm 0.10 \, \text{fm}^{ 2 } $, which supports the results of the LSSM and QVSM calculations.
Note that the third-order contribution $ a_{ n } $ does not disturb this correlation.
In fact, the LSSM calculation with the SNV effective interaction yields $ a_{ n } = 0.053 \, \text{fm}^{ 2 } $ and $ a_{ p } = - 0.001 \, \text{fm}^{ 2 } $ for $^{129}$Xe.

\subsection{$^{199}$Hg}
\label{subsec_discussion-199Hg}

No apparent correlation with magnetic moment was found for the $^{199}$Hg case.
The predicted value of the magnetic moment in the second-lowest $ 1/2^- $ state is very close to the value of the ground state.
This implies that the magnetic moment is useless for discussing the contamination in the ground-state wave function of $^{199}$Hg.


\begin{table*}[htb]
\caption{
\label{table: NSM of Hg-199}
Calculated NSM coefficients $ s_{ N } $ and $ a_{ N } $ for $^{199}$Hg in units of $ \text{fm}^{ 2 } $.
The last three columns show the contributions of the $ P $, $ T $-odd $ \pi NN $ interactions given in Table~III of Ref.~\cite{Yanase2020-129Xe-199Hg}.
The second and third rows show the LSSM calculations for the ground state and second lowest $ 1/2^- $ state, respectively.
}
\begin{center}
\begin{tabular*}{\textwidth}{@{\extracolsep{\fill}}lcccccccc} \hline
					&	$ s_{ p } $	& $ s_{ n } $ 	& $ s_{ p } + s_{ n } $
					& $ a_{ p } $ & 	$ a_{ n } $
					& $ a_{ 0 } $ 	& $ a_{ 1 } $ 	& $ a_{ 2 } $ \\\hline
	IPM 		& $ 0 $ 			& $ 2.84 $ 	& $ 2.84 $
					& $ 0 $ 			&	$ - 0.15 $
					& $ 8.4 $ 	& $ 8.4 $ 	& $ 16.7 $ \\
	LSSM (g.s.)
					& $ 0.02 $	& $ 2.44 $ 	& $ 2.46 $
					& $ - 0.0002 $ & $ - 0.14 $
					& $ 8.0 $ 	& $ 7.8 $ 	& $ 14.7 $ \\
	LSSM ($ 1/2^-_2 $)
					& $ - 0.0001 $		& $ - 1.46 $ & $ -1.46 $
					& &
					& $ -0.05 $ 	& $ 0.4 $ 	& $ 1.3 $ \\[5pt]
	RPA~\cite{Dmitriev2003-PRL,Dmitriev2004}
					& $ 0.20 $ 	& $ 1.895 $ 	& $ 2.10 $ \\
	HF (SLy4)~\cite{Ban2010}
					& & & $ 0.3 $ & &
					& $ 1.3 $ 	& $ - 0.6 $ 	& $ 2.2 $ \\
	HF (SIII)~\cite{Ban2010}
					& & & $ 0.4 $ & &
					& $ 1.2 $ 	& $ 0.5 $ 		& $ 1.6 $ \\
	HF (SV)~\cite{Ban2010}
					& & & $ 0.2 $ & &
					& $ 0.9 $ 	& $ -0.01 $ 	& $ 1.6 $ \\
	HFB (SLy4)~\cite{Ban2010}
					& & & $ 0.7 $ & &
					& $ 1.3 $ 	& $ - 0.6 $ 	& $ 2.4 $ \\
	HFB (SkM*)~\cite{Ban2010}
					& & & $ 1.3 $ & &
					& $ 4.1 $ 	& $ - 2.7 $ 	& $ 6.9 $ \\\hline
\end{tabular*}
\end{center}
\end{table*}


Table~\ref{table: NSM of Hg-199} summarizes the NSM coefficients for $^{199}$Hg.
The IPM estimate is attributed to the last neutron in the $ 2p_{ 1/2 } $ orbital, whereas the lower neutron orbitals are fully occupied.
The LSSM calculation is performed by employing the Kuo-Herling effective interaction.
We use the same empirical formula for the harmonic oscillator frequency in the case of $^{129}$Xe.
The resulting root-mean-square charge radius $ 5.3 \, \text{fm} $ is consistent with the experimental value of $ 5.4 \, \text{fm} $.

The LSSM calculation shows that the NSM coefficient $ s_{ n } $ is moderately reduced from the IPM estimate.
This result corresponds to a higher sensitivity to the neutron EDM, whereas the current experimental limit of $^{199}$Hg obtained using the RPA results is as tight as the direct measurement.
The second term in Eq.~(\ref{eq: S2 operator}) yields the main contribution $ 2.98~\text{fm}^{ 2 } $ in the ground state, whereas the contributions of the first and second terms are comparable in the second-lowest $ 1/2^- $ state.
This is qualitatively consistent with the RPA results.
However, the HF and HFB calculations in Ref.~\cite{Ban2010} yielded significantly smaller second-order NSM coefficients.
Their results, which correspond to the long-range contribution from pion exchange, can be interpreted as the summation of $ s_{ p } $ and $ s_{ n } $.
If the proton EDM contribution $ s_{ p } $ is negligible, as expected from the LSSM and RPA calculations, one can find possible correlations between $ s_{ n } $, $ a_{ 0 } $, and $ a_{ 2 } $.
As discussed in Ref.~\cite{Yanase2020-129Xe-199Hg}, these correlations might be attributed to the contamination of excited states, as in the case of $^{129}$Xe.
The ground state of $^{199}$Hg can be identified by the spectroscopic factor and collective nature of some excited states instead of the magnetic moment.

\section{Conclusions}
\label{sec_conclusions}

In summary, we computed the NSM coefficients of $^{129}$Xe and $^{199}$Hg induced by the nucleon EDM, using the nuclear shell model.
For $^{129}$Xe, we found a clear correlation between the leading-order contribution $ s_{ n } $ and the magnetic moment, which led to a significant reduction in the theoretical uncertainty.
The inclusion of the $ 0g_{ 9/2 } $ and $ 0h_{ 9/2 } $ orbitals and the next-to-leading order contribution $ a_{ n } $ do not disturb this useful correlation.
We conclude from the LSSM calculations that the sensitivity of the $^{129}$Xe NSM to the neutron EDM is given by $ s_{ n } + a_{ n } = 0.47~\text{fm}^{ 2 } $.
For $^{199}$Hg, we demonstrated that the large discrepancies in the NSM coefficients, $ s_{ n } $ and $ a_{ T } $, between the nuclear model calculations can be attributed to the contamination of excited states, as well as the case of $^{129}$Xe.
Using our results, $ s_{ n } + a_{ n } = 2.3~\text{fm}^{ 2 } $, together with the current experimental limit on the $^{199}$Hg atomic EDM~\cite{Graner2016,Graner2017-erratum} and a recent atomic many-body calculation~\cite{Sahoo2018}, one can place a tight constraint on the neutron EDM as $ \big| d_{ n } \big| < 1.8 \times 10^{-26} \, e \, \text{cm} $.

\section*{Acknowledgement}

This research was supported by MEXT as ``Program for Promoting Researches on the Supercomputer Fugaku'' (Simulation for basic science: from fundamental laws of particles to creation of nuclei) and JICFuS, and JSPS KAKENHI Grants No. 20K03925, 22K03600, and 22K14031.
This research used computational resources of supercomputer Fugaku provided by RIKEN through the HPCI System Research Project (Project ID:hp220174,  hp210165) and of Oakforest-PACS and Wisteria provided by Multidisciplinary Cooperative Research Program in Center for Computational Sciences, University of Tsukuba.
The KSHELL code was utilized for the LSSM calculations~\cite{Shimizu2019-KSHELL}.
We would like to thank Editage (www.editage.com) for English language editing.

\bibliographystyle{elsarticle-num}
\bibliography{edm,nuclear-structure}

\end{document}